\documentclass[12pt,preprint]{aastex}

\def\specchar#1{{\sc #1}}

\def\SiI{\mbox{Si\,\specchar{i}}}

\def\arcsec{\hbox{$^{\prime\prime}$}}

\def\pun{\stackrel{}{\mbox{.}}}
\def\farcs{$\stackrel{\prime\prime}{\pun}$}

\def\kms{\hbox{km$\;$s$^{-1}$}}
\def\ms{\hbox{m$\;$s$^{-1}$}}

\shorttitle{Channeling the 5-min oscillations into the chromosphere}

\shortauthors{Khomenko et al.}

\begin{document}

\title{Channeling 5-min photospheric oscillations into the solar outer atmosphere
through small-scale vertical magnetic flux tubes}

\author{E. Khomenko\altaffilmark{1,2}, R. Centeno\altaffilmark{3},
        M. Collados\altaffilmark{1} and J. Trujillo Bueno\altaffilmark{1,4} }
\email{khomenko@iac.es, rce@ucar.edu, mcv@iac.es, jtb@iac.es}

\altaffiltext{1}{Instituto de Astrof\'{\i}sica de Canarias, 38205 La Laguna, Tenerife, Spain}
\altaffiltext{2}{Main Astronomical Observatory, NAS, 03680, Kyiv,
Ukraine} \altaffiltext{3}{High Altitude Observatory, National
Center for Atmospheric Research\footnote{The National Center for
Atmospheric Research is sponsored by the National Science
Foundation.}, Boulder, CO 80301.}
\altaffiltext{4}{Consejo Superior de Investigaciones Cient\'\i ficas, (Spain)}

\begin{abstract}
We report two-dimensional MHD simulations which demonstrate
that photospheric 5-min oscillations can leak into the
chromosphere inside small-scale vertical magnetic flux tubes.
The results of our numerical experiments are compatible with those
inferred from simultaneous spectropolarimetric observations of the
photosphere and chromosphere obtained with the Tenerife Infrared
Polarimeter (TIP) at 10830 \AA. We conclude that the efficiency of
energy exchange by radiation in the solar photosphere can lead to
a significant reduction of the cut-off frequency and may allow for
the propagation of the 5 minutes waves vertically into the
chromosphere.
\end{abstract}



\section{Introduction}

Several recent investigations have reported the detection of 5
minute oscillations in the chromosphere and corona of the Sun
\citep{Krijer+etal2001, DeMoortel+etal2002, DePontieu+etal2003, Centeno+etal2006b,
Veccio+etal2007}. These 5-min oscillations
have been detected mainly above facular and network areas, while
oscillations in the outer atmospheric regions of sunspots show
mainly a 3-min periodicity \citep[][and references
therein]{Centeno+etal2006a, Bogdan+Judge2006}, similar to what it is
found in the internetwork regions of the Sun.
An explanation for the shift in the wave period with height from 5
to 3 minutes was suggested by \citet{Fleck+Schmitz1991}, who argue
that this is a basic phenomenon due to resonant excitation at the
atmospheric cut-off frequency (see also Fleck \& Schmitz 1993).
The low temperatures of the high
photosphere give rise to the 3-minute cut-off period.
However, it is unclear why such a shift in the wave period is not
observed in small-scale structures present in facular and network
regions.
How can the evanescent 5 minute oscillations propagate up to the
chromospheric heights?

\citet{DePontieu+etal2004} argue that the inclination of the
magnetic field is essential for the leakage of p-modes strong
enough to produce the dynamic jets observed in active region
fibrils. Following the same concept,
\citet{Jefferies+etal2006} claim that inclined flux tubes explain
their lower chromospheric observations of propagating waves
(whatever their amplitude).
If (acoustic or slow MHD) waves have a preferred direction of
propagation defined by the magnetic field, the effective cut-off
frequency is lowered by the cosine of the inclination angle with
respect to the solar local vertical. This allows evanescent waves
to propagate.
However, there are also works reporting vertically
propagating 5 minute waves in facular and network regions at
chromospheric heights that cannot be easily explained by the
above-mentioned mechanism \citep[see e.g.,][and references
therein]{Krijer+etal2001, Centeno+etal2006b, Veccio+etal2007}.
Alternatively, a decrease in the effective acoustic cut-off
frequency can be produced by taking into account the radiative
energy losses, without the need of assuming inclined magnetic flux
tubes \citep{Roberts1983, Centeno+etal2006b}. If the radiative
relaxation time ($\tau_{RR}$) is sufficiently small (as expected
for small-scale magnetic structures), the cut-off frequency can be
reduced allowing the 5-min evanescent waves to propagate.

In this letter, we consider the observational evidences provided
by \citet{Centeno+etal2006b}, and extend the theoretical analysis
by \citet{Roberts1983} with the help of non-adiabatic, non-linear
2D numerical simulations of magneto-acoustic waves in small-scale
flux tubes with a realistic magnetic field configuration.

\section{Summary of the spectropolarimetric observations}

A facular region, located close to the disc center ($\mu=$ 0.95),
was observed at the German Vacuum Tower Telescope (VTT) of the
Observatorio del Teide on 14 July 2002, using the Tenerife
Infrared Polarimeter \citep{MartinezPillet+etal1999}. The data set
consists of a time series of 81 minutes (with a temporal cadence of 5.4
s) taken with a fixed slit position.
The slit was 0\farcs5 wide and 40\arcsec\ long.
The full Stokes profiles were measured in a spectral range that
spanned from 10825.5 to 10833 \AA, with a spectral sampling of 31
m\AA\ per pixel. This spectral region includes a photospheric
\SiI\ line at 10827.09 \AA\ and a chromospheric Helium {\sc i}
10830 \AA\ line, which is indeed a triplet whose blue component
($\lambda$ 10829.09\AA) is quite weak and difficult to see in an
intensity spectrum, and whose red components ($\lambda$ 10830.25,
$\lambda$ 10830.34 \AA) appear blended. Several non-LTE radiative
transfer investigations indicate that the He {\sc i} 10830 \AA\
spectral line radiation is originated in a relatively thin layer
in the upper chromosphere, about 2000 km above the base of the
photosphere \citep{Pozhalova1988, Avrett+etal1994,
Centeno+etal2008a}.
The LOS velocity variations of these spectral lines have been the
main object of the analysis. They were retrieved by applying a
Milne-Eddington (ME) inversion to the chromospheric Helium lines
and a standard LTE inversion code to the photospheric Silicon line.
For a detailed description of the observations and their analysis,
see \citet{Centeno+etal2006b, Centeno+etal2008b}. Below we
summarize briefly the main findings.

\begin{figure*}
\epsscale{1.0}
\plottwo{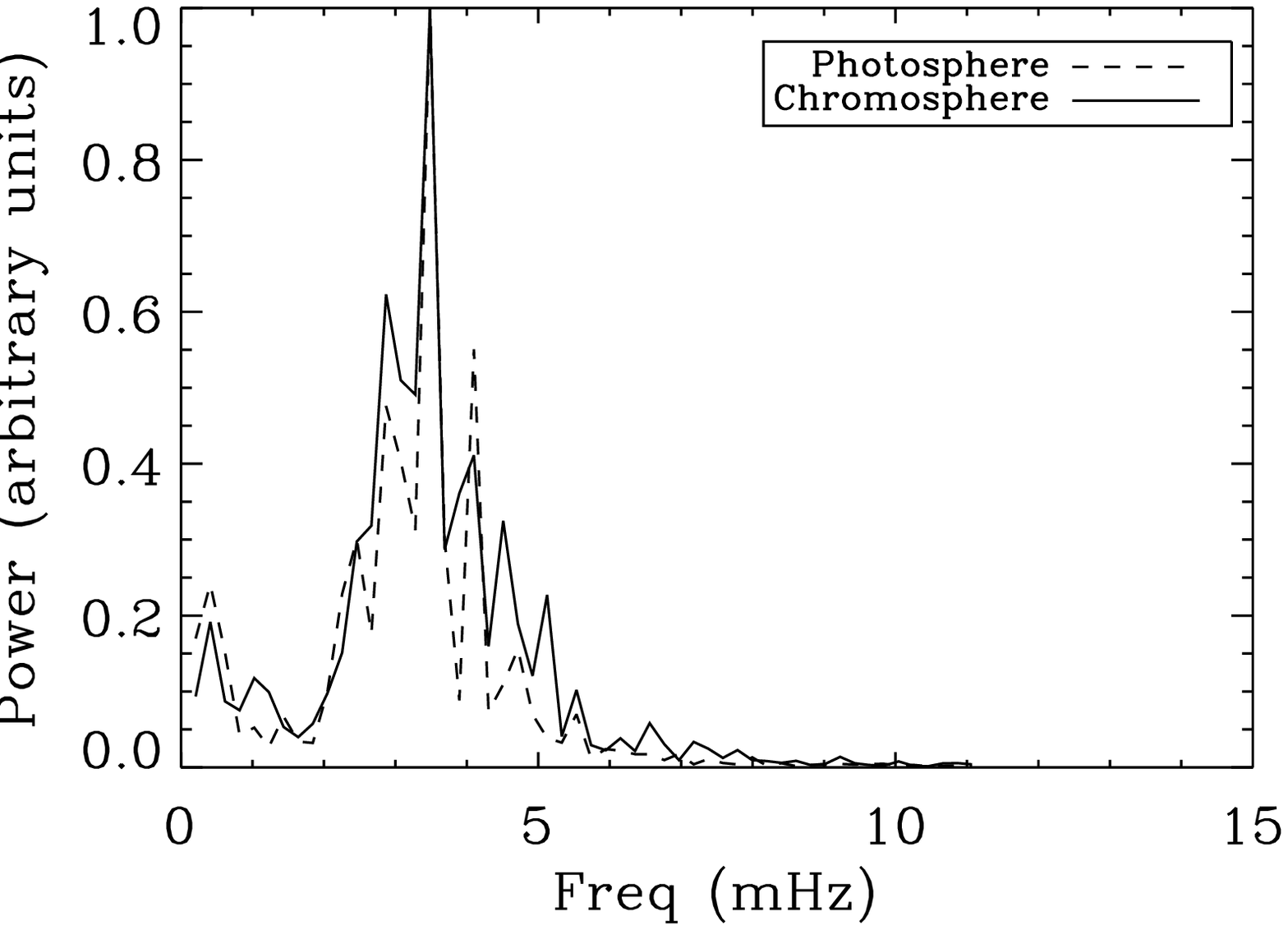}{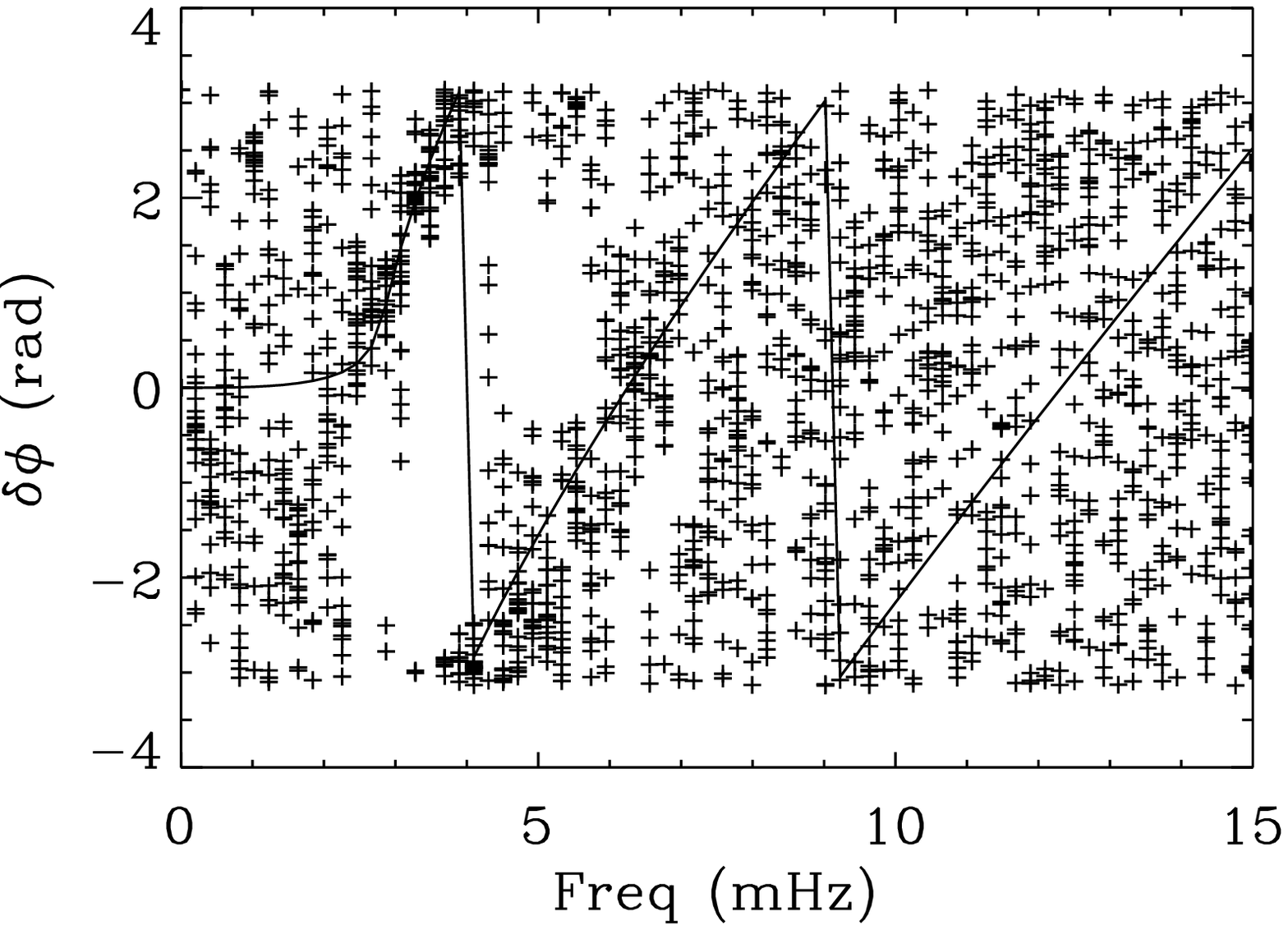}
\caption{Left: Photospheric (solid line) and chromospheric (dashed
line) LOS velocity power spectra in the observed plage region.
Right: Phase spectrum for the LOS velocity oscillations.  Each
cross on the figure is computed as the difference from the
chromospheric phase and the photospheric phase for one frequency
and one position along the slit. Solid lines represent the best
fit of the theoretical model with values T=9500 K, $\tau_{RR}=10$
s, and $\delta z$ = 1500 km. \label{fig:obs}}
\end{figure*}

The left panel of Fig.~\ref{fig:obs} shows the power spectra of
velocity averaged over all the positions along the slit in the
photosphere (dashed line) and the chromosphere (solid line) as a
function of the frequency of the oscillating mode. The right panel
represents the difference in phase ($\delta\phi$)
between the chromospheric and the photospheric velocity
oscillations, also as a function of frequency.
It is interesting to note that the photospheric and chromospheric
facular power spectra show very similar features with peaks at the
same positions, having their main contribution in the 3 mHz band
(5-minute oscillations).
The facular phase difference spectrum shows a very noisy behavior
below 2 mHz, indicating that there is no wave propagation in this
frequency regime. Above this point (and also around the 3 mHz
band) the phase difference starts to increase with frequency,
meaning that these frequency modes do propagate from the
photosphere, reaching the chromosphere some time later.

\section{Theoretical considerations}

\begin{figure}
\epsscale{0.5} \plotone{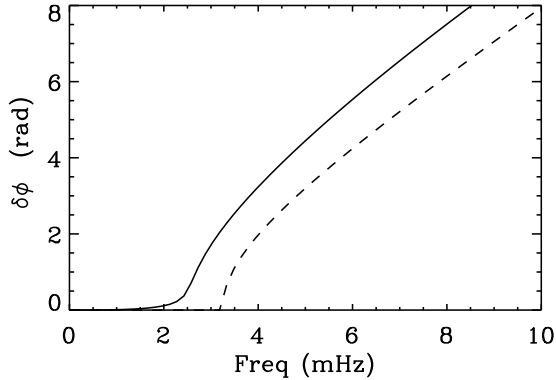} \caption{Phase difference
between waves at different heights calculated theoretically as a
function of frequency. Dashed line: assuming
adiabatic oscillations; solid
line: using a radiative relaxation time of $\tau_{RR}=10$ s.
Note the difference in the effective cut-off frequency between the
two cases. The parameters of the calculation are $\delta z $ =
1500 km, $T=$9000 K. \label{fig:phase}}
\end{figure}

The low amplitudes of the observed linear polarization signals of
the \SiI\ line (not larger than $2 \times 10^{-3}$ in units of the
continuum intensity) imply inclination angles smaller than 10-15
degrees to the solar vertical.
Consequently, throughout the present paper we make the assumption
that the magnetic field observed in the facular region is close to
the local vertical.
The equations describing the vertical propagation of a wave in a
vertical magnetic field in a gravitationally stratified isothermal
atmosphere are formally equal to those governing the propagation
of an acoustic wave \citep{Ferraro+Plumpton1958, Roberts2006}.
These equations have the following well known solution describing the
variation of the wave amplitude $V(z)$ with height:
\begin{equation}
V(z)=V_0 e^{z/2H} e^{k_z z} \,,
\end{equation}
where $k_z^2 = ( \omega^2 - \omega_*^2) / c^2_*$, $c^2_*=\gamma_* g H$,
$\omega_*^2=\gamma_* g/4 H$ and $\gamma_*=(1 - i \gamma \omega \tau_{RR})/(1 - i \omega
\tau_{RR})$ \citep{Mihalas+Mihalas1984}.
In general, the vertical wave vector $k_z$ is a complex number. If
the wave propagation is adiabatic ($\tau_{RR} \rightarrow \infty
$), the real part of the wave vector $k_r$ becomes equal to zero
for frequencies below the cut-off and no propagation is possible.
However, if energy exchange by radiation is taken into
account --that is, if the radiative relaxation time
$\tau_{RR}$ has a finite value \citep[see][]{Spiegel1957}, $k_z$ always
has both an imaginary and a real part. Thus, formally, the wave
propagation is possible at all frequencies in this case.
It should be noted that
the radiative relaxation time is very small in small-scale structures
of the solar photosphere \citep[see][]{Kneer+TrujilloBueno1987}.
The properties of the wave solution are illustrated in
Fig.~\ref{fig:phase}. It shows the phase difference $\delta \phi =
k_r \delta z$ between the waves at two heights (separated 1500 km)
as a function of frequency for the cases $\tau_{RR} \rightarrow
\infty$ (dashed line) and $\tau_{RR}$=10 s (solid line). An
effective cut-off frequency may be defined for the case of
$\tau_{RR}$=10 s, being significantly lower than for the
adiabatic case. This fact causes important effects on wave
propagation.

The solid curve on the right panel of Fig.~\ref{fig:obs} gives the
best fit of the above-mentioned model to the observed phase
spectrum in the facular region. The free parameters of the model
are the temperature, the radiative relaxation time and the height
difference between the two velocity signals. The best fit results in
a value of $\tau_{RR}=10$ s.

\begin{figure}
\epsscale{0.5} \plotone{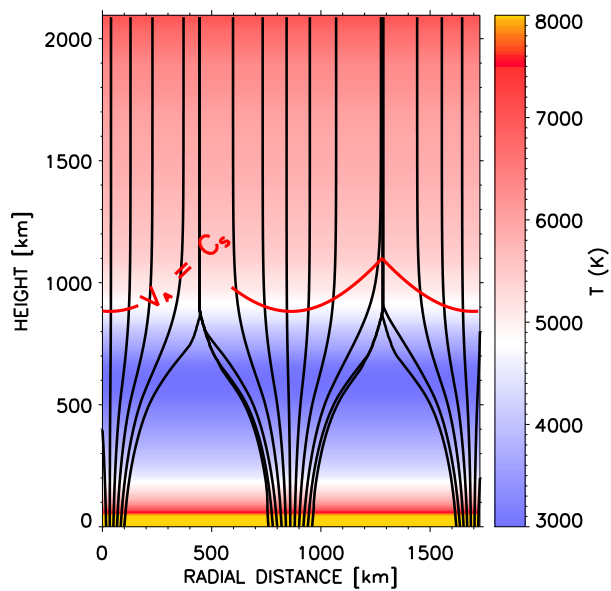} \caption{Magnetostatic flux tube
model. The background image quantifies the temperature values. The
magnetic field along the flux tube axis varies with height from
740 G in the photosphere to 37 G at z=2000 km, the plasma $\beta$
decreases from 4 to 7$\times 10^{-4}$ and the cut-off period
increases from 230 s in the photosphere to 280 s in the
chromosphere. \label{fig:tube}}
\end{figure}

\section{MHD simulations}

To generalize these conclusions (based on a simple model of linear
wave propagation) to the case of a more realistic atmosphere with
horizontal and vertical stratification in all the physical
parameters, a complex magnetic field configuration and non-linear
wave propagation, we solved numerically the set of MHD equations
in this realistic environment.
The numerical code used for the solution of these equations is
described elsewhere \citep{Khomenko+Collados2006,
Khomenko+Collados2007}. Radiative losses were taken into account
by means of Newton's law of cooling.
The initial magnetostatic flux tube model was constructed
following the method described by \citet{Pneuman+etal1986} (see
Fig.~\ref{fig:tube}).
%

Previous investigations showed that the photospheric driver that
excites oscillations inside the flux tube should have a vertical
component in order to generate an acoustic fast wave already in
the photosphere \citep{Khomenko+Collados2007}. If this were not the
case, no correlation would be observed between the photospheric and
chromospheric velocity signals, contrarily to what is suggested by
observations (see the power spectra of Fig.~\ref{fig:obs}).

Thus, we applied a vertical periodic driver at the bottom boundary
of the simulation domain by varying the vertical velocity as:
$V_z(x,\it{t})=V_0 \sin(\omega \it{t})\times
\exp(-(x-x_0)^2/2\sigma^2)$, where $\sigma=160$ km is the
horizontal size of the pulse, $V_0=200$ \ms\ is the initial
amplitude at the photosphere and $x_0$ corresponds to the position
of the tube axis.
The period of the driver was 300 s (i.e., it is above the cut off period).
In what follows we will compare two simulation runs that are
identical except for the value of the radiative relaxation time.
While the first run is in the adiabatic regime ($\tau_{RR}
\rightarrow \infty$), the second one was carried out with  $\tau_{RR}=10$ s
(constant through the whole atmosphere).

\begin{figure*}
\epsscale{1.0} \plotone{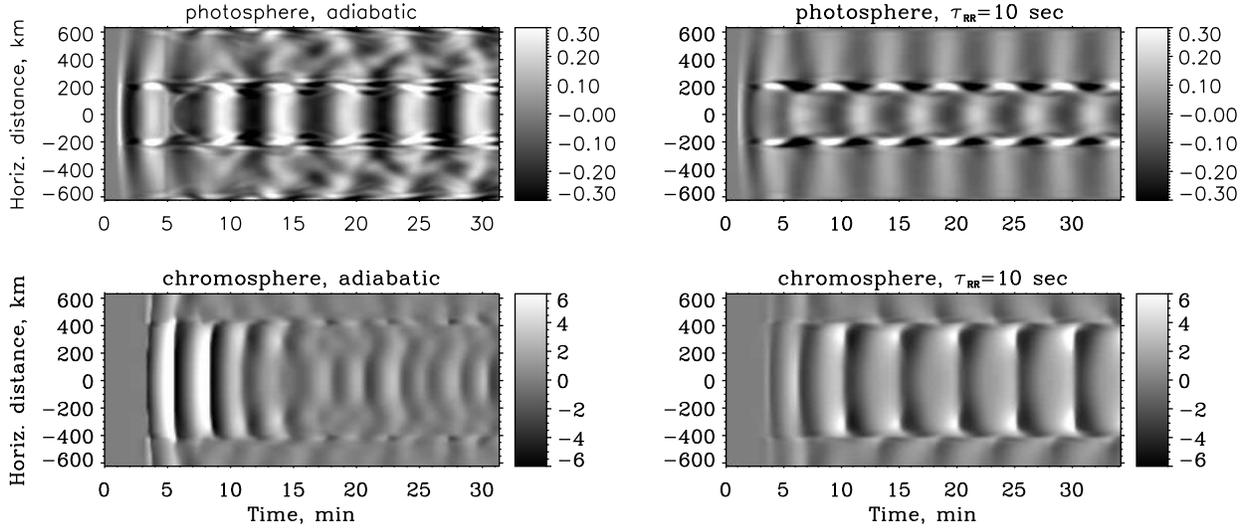} \caption{Time-distance
plots of the vertical velocity from the simulations at
photospheric (top) and chromospheric (bottom) heights. Left:
adiabatic calculation; right: calculation with a radiative
relaxation time $\tau_{RR}=10$ s. Note how the horizontal extent
of the tube is larger in the chromosphere because of the field
expansion. The vertical bars on the right of each plot represent
the velocity scale in \kms. \label{fig:td}}
\end{figure*}

\begin{figure*}
\epsscale{1.0} \plotone{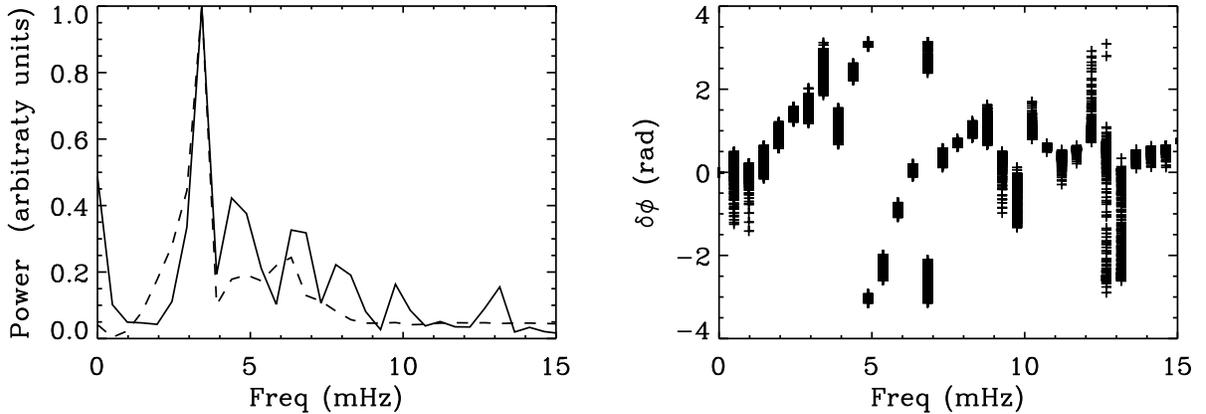} \caption{Left:
photospheric (dashed line) and chromospheric (solid line) power
spectra obtained from the simulation with $\tau_{RR}$ equal to 10
s. Right: phase difference spectrum for the vertical velocity
oscillations. Each cross on the figure is computed as the
difference from the chromospheric phase and the photospheric phase
for one frequency and one position inside the flux tube.
\label{fig:spectras}}
\end{figure*}

Fig.~\ref{fig:td} gives time-distance plots of the vertical
velocity obtained from the simulations at two heights: the upper
photosphere around 400 km (top) and the chromosphere around 1500
km (bottom).
The vertical photospheric driver generates a fast magneto-acoustic
wave that is essentially acoustic in the photosphere where $\beta
> 1$. This wave propagates upwards through the $\beta=1$ transition
layer, preserving its acoustic nature and being transformed
into a slow magneto-acoustic mode higher up.

In the adiabatic case (left panels of Fig. 4), we
can see that the velocity in the photosphere has a well defined
5-minute periodicity within the flux tube (horizontal distance
from -200 to 200 km). Higher up in the chromosphere, this behavior
changes. Note that it takes some time for the oscillations to
reach the chromospheric heights (bottom panels of
Fig.~\ref{fig:td}). In the chromosphere the tube expands and
occupies now a larger area in the time-distance plot (from about
-400 to 400 km). At the beginning of the time series, two strong
shock waves develop, and are followed in time by strongly damped
oscillations. Only weak shocks exist in the stationary phase of
the simulations and the wave periodicity is 3 minutes. Thus, in
the adiabatic case, there is a shift with height of the dominant
period from five minutes in the photosphere to three minutes in
the chromosphere.

This behavior contrasts with that of the simulations with
$\tau_{RR}=10$ s (right panels of Fig.~\ref{fig:td}). Contrarily
to what happens in the adiabatic case, the 5-minute periodicity is
preserved both in the photosphere and in the chromosphere. Waves
are linear in the deeper layers, they propagate vertically upwards
through the temperature minimum and form shocks with 5-6 \kms\
amplitudes in the higher layers.

The left panel of Fig.~\ref{fig:spectras} shows the power spectra
of the photospheric and chromospheric vertical velocity
oscillations in the simulation run with $\tau_{RR}=10$ s.
Now there is no shift in the wave period as we go to the
higher atmospheric layers, and both
the photospheric and chromospheric power spectra are rather similar
(right panel). This figure can be directly compared to the
observed power spectra in Fig.~\ref{fig:obs} (left panel). The
similarity between the observed and the simulated power spectra is
notable.
The right panel of Fig.~\ref{fig:spectras} gives the difference in
phase between the chromospheric and the photospheric velocity
oscillations as a function of frequency, as calculated from the
simulations. This figure can be compared with the observed phase
spectra in Fig.~\ref{fig:obs} (right panel). It can be seen that
the simulations describe correctly the phase speed of waves in the
frequency range 0--5 mHz, as well as the value of the effective
cut-off frequency. This gives additional support to our
conclusions.
Radiative losses are expected to play an important role in
small-scale magnetic structures, such as those present in facular
regions and, as already shown, they are able to decrease the
cut-off frequency allowing the five-minute oscillation to
propagate into the chromosphere.


The results shown above were obtained with a fixed value of the
radiative relaxation time ($\tau_{RR}=10$ s).  We have checked
that similar conclusions are reached with a height-dependent
relaxation time changing from 10 s in the photosphere to 100 s in
the chromosphere.

\section{Conclusions}

In this paper we have performed an observational, analytical and
numerical analysis of the propagation of the 5-min oscillations from
the photosphere to the chromosphere in small-scale vertical magnetic flux tubes.
From the observational point of view, it has been demonstrated
that the power spectra of oscillations in the photosphere and the
chromosphere in a facular region are similar and peak at 3 mHz.
The observed velocity variations are compatible with the
longitudinal acoustic wave propagation in a vertical magnetic
field from the photosphere to the chromosphere.
The analytical theory of acoustic waves in a stratified isothermal
atmosphere embedded in a vertical magnetic field suggests that the
effective cut-off frequency can be significantly reduced if the
radiative losses of the oscillations are taken into account. This
allows the propagation of the evanescent waves at 3 mHz.

The numerical solution of the complete set of the MHD equations
allows to generalize this conclusion to the situation of a
complex magnetic field configuration, stratification in all
atmospheric parameters and non-linear wave effects.
We conclude that our photospheric and chromospheric
observations can be explained assuming that the radiative relaxation
time is sufficiently small (as expected in small-scale magnetic
structures). This significantly reduces the cut-off frequency and
makes evanescent 5-minute waves propagate vertically from the
photosphere towards the chromosphere without changing their period
with height.
Our conclusion that vertical magnetic field concentrations can
also channel the 5-min oscillatory power of the photosphere
towards the chromosphere in facular regions extends the
actual possibilities of the dynamic and magnetic couplings between
these two important layers of the solar atmosphere.
Additional work on three-dimensional MHD simulations with
non-LTE radiative transfer physics will be necessary to
investigate the role of this mechanism in more realistic models of
the highly dynamic conditions of the magnetized solar atmosphere.

\acknowledgements  The authors are grateful to C. Beck for
helpful discussions. This research has been funded by the Spanish
Ministerio de Educaci{\'o}n y Ciencia through projects
AYA2007-63881 and AYA2007-66502.



\begin{thebibliography}{30}
\expandafter\ifx\csname natexlab\endcsname\relax\def\natexlab#1{#1}\fi

\bibitem[{Avrett {et~al.}(1994)Avrett, Fontenla, \& Loeser}]{Avrett+etal1994}
Avrett, E.~H., Fontenla, J.~M., \& Loeser, R. 1994, in Infrared Solar Physics,
  IAU Symp. No. 154, ed. D.~Rabin, J.~Jefferies, \& C.~Lindsey, Vol. 154,
  35---47

\bibitem[{Bogdan \& Judge(2006)}]{Bogdan+Judge2006}
Bogdan, T.~J. \& Judge, P.~G. 2006, in MHD wave and oscillations in the Solar
  Plasma, Vol. 364, Issue 1839 (Phil. Trans. Royal. Soc.), 313---331

\bibitem[{Centeno {et~al.}(2006{\natexlab{a}})Centeno, Collados, \&
  \mbox{Trujillo Bueno}}]{Centeno+etal2006b}
Centeno, R., Collados, M., \& \mbox{Trujillo Bueno}, J. 2006{\natexlab{a}}, in
  Solar Polarization 4, ed. R.~Casini \& B.~W. Lites, Vol. 358, ASP Conference
  Series, 465

\bibitem[{Centeno {et~al.}(2006{\natexlab{b}})Centeno, Collados, \&
  \mbox{Trujillo Bueno}}]{Centeno+etal2006a}
Centeno, R., Collados, M., \& \mbox{Trujillo Bueno}, J. 2006{\natexlab{b}},
  ApJ, 640, 1153

\bibitem[{Centeno {et~al.}(2008{\natexlab{a}})Centeno, Collados, \&
  \mbox{Trujillo Bueno}}]{Centeno+etal2008b}
---. 2008{\natexlab{a}}, ApJ, in preparation

\bibitem[{Centeno {et~al.}(2008{\natexlab{b}})Centeno, \mbox{Trujillo Bueno},
  Uitenbroek, \& Collados}]{Centeno+etal2008a}
Centeno, R., \mbox{Trujillo Bueno}, J., Uitenbroek, H., \& Collados, M.
  2008{\natexlab{b}}, ApJ, submitted

\bibitem[{\mbox{De Moortel} {et~al.}(2002)\mbox{De Moortel}, Ireland, Hood, \&
  Walsh}]{DeMoortel+etal2002}
\mbox{De Moortel}, I., Ireland, J., Hood, A.~W., \& Walsh, R.~W. 2002, A\&A,
  387, L13

\bibitem[{\mbox{De Pontieu} {et~al.}(2003)\mbox{De Pontieu}, Erdelyi, \&
  de~Wijn}]{DePontieu+etal2003}
\mbox{De Pontieu}, B., Erdelyi, R., \& de~Wijn, A.~G. 2003, ApJ, 595, L63

\bibitem[{\mbox{De Pontieu} {et~al.}(2004)\mbox{De Pontieu}, Erdelyi, \&
  Stewart}]{DePontieu+etal2004}
\mbox{De Pontieu}, B., Erdelyi, R.~J., \& Stewart, P. 2004, Nat, 430, Issue
  6999, 536

\bibitem[{Ferraro \& Plumpton(1958)}]{Ferraro+Plumpton1958}
Ferraro, V. C.~A. \& Plumpton, C. 1958, ApJ, 127, 459

\bibitem[{Fleck \& Schmitz(1991)}]{Fleck+Schmitz1991}
Fleck, B. \& Schmitz, F. 1991, A\&A, 250, 235

\bibitem[{Fleck \& Schmitz(1993)}]{Fleck+Schmitz1993}
---. 1993, A\&A, 273, 671

\bibitem[{Jefferies {et~al.}(2006)Jefferies, McIntosh, Armstrong, Bogdan,
  Thomas, Cacciani, \& Fleck}]{Jefferies+etal2006}
Jefferies, S.~M., McIntosh, S.~W., Armstrong, J.~D., Bogdan, T., Thomas, J.,
  Cacciani, A., \& Fleck, B. 2006, ApJ, 648, L151

\bibitem[{Khomenko \& Collados(2006)}]{Khomenko+Collados2006}
Khomenko, E. \& Collados, M. 2006, ApJ, 653, 739

\bibitem[{Khomenko \& Collados(2007)}]{Khomenko+Collados2007}
---. 2007, Solar Phys., submitted

\bibitem[{Kneer \& \mbox{Trujillo-Bueno}(1987)}]{Kneer+TrujilloBueno1987}
Kneer, F. \& \mbox{Trujillo Bueno}, J. 1987, A\&A, 183, 91

\bibitem[{Krijger {et~al.}(2001)Krijger, Rutten, Lites, Straus, Shine, \&
  Tarbell}]{Krijer+etal2001}
Krijger, J.~M., Rutten, R.~J., Lites, B.~W., Straus, T., Shine, R.~A., \&
  Tarbell, T.~D. 2001, A\&A, 379, 1052


\bibitem[{\mbox{Mart\'{\i}nez Pillet} {et~al.}(1999)\mbox{Mart\'{\i}nez Pillet},
  {Collados}, \mbox{S{\'a}nchez Almeida}, {Gonz{\'a}lez}, {Cruz-Lopez},
  {Manescau}, {Joven}, {Paez}, {Diaz}, {Feeney}, {S{\'a}nchez}, {Scharmer}, \&
  {Soltau}}]{MartinezPillet+etal1999}
\mbox{Mart\'{\i}nez Pillet}, V., {Collados}, M., \mbox{S{\'a}nchez Almeida}, J.,
  {Gonz{\'a}lez}, V., {Cruz-Lopez}, A., {Manescau}, A., {Joven}, E., {Paez},
  E., {Diaz}, J., {Feeney}, O., {S{\'a}nchez}, V., {Scharmer}, G., \& {Soltau},
  D. 1999, in High resolution solar physics: theory, observations and
  techniques, ed. T.~R. Rimmele, K.~S. Balasubramaniam, \& R.~R. Radick, Vol.
  183, 19th NSO/SP Summer Workshop (ASP Conf. Series), 264

\bibitem[{Mihalas \& Mihalas(1984)}]{Mihalas+Mihalas1984}
Mihalas, D. \& Mihalas, B.~W. 1984, Foundations of Radiation Hydrodynamics
  (Oxford: Oxford University Press)


\bibitem[{Pneuman {et~al.}(1986)Pneuman, Solanki, \&
  Stenflo}]{Pneuman+etal1986}
Pneuman, G.~W., Solanki, S.~K., \& Stenflo, J.~O. 1986, A\&A, 154, 231

\bibitem[{Pozhalova(1988)}]{Pozhalova1988}
Pozhalova, Z.~A. 1988, Soviet Astr., 32, N 5, 542


\bibitem[{Roberts(1983)}]{Roberts1983}
Roberts, B. 1986, Solar Phys., 87, 77

\bibitem[{Roberts(2006)}]{Roberts2006}
Roberts, B. 2006, Royal Society of London Transactions Series A, 364, Issue
  1839, 447


\bibitem[{Spiegel(1957)}]{Spiegel1957}
Spiegel, E.~A. 1957, ApJ, 126, 202

\bibitem[{Vecchio {et~al.}(2007)Vecchio, Cauzzi, Reardon, Janssen, \&
  Rimmele}]{Veccio+etal2007}
Vecchio, A., Cauzzi, G., Reardon, K.~P., Janssen, K., \& Rimmele, T. 2007,
  A\&A, 461, L1


\end{thebibliography}
\end{document}